\documentclass[twocolumn,english,aps,prl,superscriptaddress]{revtex4}

\usepackage{times}

\usepackage{graphicx}
\usepackage{babel}
\usepackage{amsmath}
\usepackage{amsfonts}
\usepackage{amssymb}
\usepackage{epsfig}

\begin{document}
\normalsize{\textbf{Science, Vol.313, p.499, 2006}}\\\\
\title{Violation of Kirchhoff's
Laws for a Coherent RC Circuit}

\author{J. Gabelli}\affiliation{Laboratoire Pierre Aigrain, D{\'e}partement
de Physique de l'Ecole Normale Sup\'erieure, 24 rue Lhomond, 75231
Paris Cedex 05, France} \author{G. F\`{e}ve}\affiliation{Laboratoire
Pierre Aigrain, D{\'e}partement de Physique de l'Ecole Normale
Sup\'erieure, 24 rue Lhomond, 75231 Paris Cedex 05,
France}\author{J.-M. Berroir}\affiliation{Laboratoire Pierre
Aigrain, D{\'e}partement de Physique de l'Ecole Normale
Sup\'erieure, 24 rue Lhomond, 75231 Paris Cedex 05,
France}\author{B. Pla\c{c}ais}\affiliation{Laboratoire Pierre Aigrain,
D{\'e}partement de Physique de l'Ecole Normale Sup\'erieure, 24 rue
Lhomond, 75231 Paris Cedex 05, France}\author{A.
Cavanna}\affiliation{Laboratoire de Photonique et Nanostructures,
UPR20 CNRS, Route de Nozay, 91460 Marcoussis Cedex,
France}\author{B. Etienne}\affiliation{Laboratoire de Photonique et
Nanostructures, UPR20 CNRS, Route de Nozay, 91460 Marcoussis Cedex,
France}\author{Y.Jin}\affiliation{Laboratoire de Photonique et
Nanostructures, UPR20 CNRS, Route de Nozay, 91460 Marcoussis Cedex,
France}\author{D.C. Glattli} \email{glattli@lpa.ens.fr}
\affiliation{Laboratoire Pierre Aigrain, D{\'e}partement de Physique
de l'Ecole Normale Sup\'erieure, 24 rue Lhomond, 75231 Paris Cedex
05, France} \affiliation{Service de Physique de l'Etat Condens{\'e},
CEA Saclay, F-91191 Gif-sur-Yvette, France}

\begin{abstract}
What is the complex impedance of a fully coherent quantum
resistance-capacitance (RC)circuit at GHz frequencies in which a
resistor and a capacitor are connected in series? While Kirchhoff's
laws predict addition of capacitor and resistor impedances, we
report on observation of a different behavior. The resistance,here
associated with charge relaxation, differs from the usual transport
resistance given by the Landauer formula. In particular, for a
single mode conductor, the charge relaxation resistance is half the
resistance quantum, regardless of the transmission of the mode. The
new mesoscopic effect reported here is relevant for the dynamical
regime of all quantum devices.
\end{abstract}

\maketitle
For a classical circuit, Kirchhoff's laws prescribe the
addition
 of resistances in series. Its failure has been a central issue in developing our understanding
 of electronic transport in mesoscopic conductors. Indeed, coherent multiple electronic
reflections between scatterers in the conductor were
 found to make the conductance non-local \cite{Landauer57,Landauer70}. A new composition law of
 individual scatterer contribution to resistance was found that led to
 the solution of the problem of electron localization \cite{Anderson81} and, later,
 to formulation of the electronic conduction in terms of
 scattering of electronic waves \cite{ButtImLandauer83}. Nonadditivity of series
 resistances, or of parallel conductances, nonlocal effects and
 negative four-point resistances \cite{GaoBO} have been
 observed in a series of transport experiments at low temperature, where phase
 coherence extends over the mesoscopic scale \cite{Review1,Review2}.
 It is generally accepted that the conductance of a phase-coherent quantum
 conductor is given by the Landauer formula and its generalization to multi-lead conductors \cite{ButtLand},
which relate the conductance to the transmission of electronic waves
by the conductance quantum $e^2/h$. But, how far is this description
robust at finite frequency where conductance combines with
nondissipative circuit elements such as capacitors or inductors ?
Are there significant departures from the dc result ? The question
is important, as recent advances in quantum information highlight
the need for fast manipulation of quantum systems, in particular
quantum conductors. High frequency quantum transport has been
theoretically addressed, showing that a quantum RC circuit displays
discrepancies with its classical counterpart
\cite{Buttiker93,Pretre96}.
\begin{figure}
\centerline{\includegraphics[width=6 cm,
keepaspectratio]{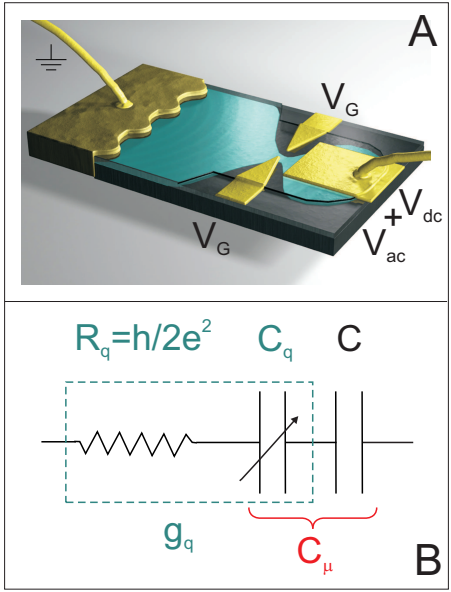}}\caption{{\small The quantum capacitor
realized using a 2DEG (A) and its equivalent circuit (B). The
capacitor consists of a metallic electrode (in gold) on top of a
submicrometer 2DEG quantum dot (in blue) defining the second
electrode. The resistor is a QPC linking the dot to a wide 2DEG
reservoir (in blue), itself connected to a metallic contact (dark
gold). The QPC voltage $V_G$ controls the number of electronic modes
and their transmission. The radio frequency voltage $V_{ac}$, and
eventually a dc voltage $V_{dc}$, are applied to the
counter-electrode whereas the ac current, from which the complex
conductance is deduced, is collected at the ohmic contact. As
predicted by theory, the relaxation resistance $R_q$, which enters
the equivalent circuit for the coherent conductance, is
transmission-independent and equal to half the resistance quantum.
The capacitance is the serial combination $C_\mu$ of the quantum and
the geometrical capacitances ($C_q$ and $C$ respectively). $C_q$ is
transmission-dependent and strongly modulated by $V_{dc}$ and/or
$V_G$. The combination of $R_q$ and $C_q$ forms the impedance
$1/g_q$ of the coherent quantum conductor.}}\label{HBT1}\end{figure}
It was shown that a counter-intuitive modification of the series
resistance lead to the situation in which the resistance is no
longer described by the Landauer formula and does not depend on
transmission in a direct way \cite{Buttiker93,Pretre96}. Instead it
is directly related to the dwell time of electrons in the capacitor.
Moreover, when the resistor transmits in a single electronic mode, a
constant resistance was found, equal to the half-resistance quantum
$h/2e^2$, i.e., it was not transmission-dependent. This resistance,
modified by the presence of the coherent capacitor, was termed a
"charge-relaxation resistance" to distinguish it from the usual dc
resistance, which is sandwiched between macroscopic reservoirs and
described by the Landauer formula. The quantum charge-relaxation
resistance, as well as its generalization in nonequilibrium systems,
is an important concept that can be applied to quantum information.
For example, it enters into the problem of quantum-limited detection
of charge qubits \cite{Pilgram} \cite{Clerk}, in the study of
high-frequency-charge quantum noise
\cite{ChargeNoise1,ChargeNoise2,ChargeNoise3}, or in the study of
dephasing of an electronic quantum interferometer \cite{Seelig}. In
molecular electronics, the charge relaxation resistance is also
relevant to the THz frequency response of systems such as carbon
nanotubes\cite{Burke}.

We report on the observation and quantitative measurement of the
quantum charge-relaxation resistance in a coherent RC circuit
realized in a two-dimensional electron gas (2DEG) (see Fig.1A). The
capacitor is made of a macroscopic metallic electrode on top of a
2DEG submicrometer dot defining the second electrode. The resistor
is a quantum point contact (QPC) connecting the dot to a wide 2DEG
macroscopic reservoir.  We address the coherent regime in which
electrons emitted from the reservoir to the dot are backscattered
without loss of coherence. In this regime, we have checked the
prediction made in refs.\cite{Buttiker93,Pretre96} that the
charge-relaxation resistance is not given by the Landauer formula
resistance but instead is constant and equals $h/2e^2$, as the QPC
transmission is varied. Note that we consider here a spin-polarized
regime and that the factor $1/2$ is not the effect of spin, but a
hallmark of a charge-relaxation resistance. When coherence is washed
out by thermal broadening, the more conventional regime pertaining
to dc transport is recovered. The present work differs from previous
capacitance measurements where, for spectroscopic purpose, the dot
reservoir coupling was weak and the ac transport regime was
incoherent \cite{Ashoori92a,Ashoori92b}. As a consequence, although
quantum effects in the capacitance were observable, the quantum
charge-relaxation resistance was not accessible in these earlier
experiments.

At zero temperature in the coherent regime and when a single mode is
transmitted by the QPC, the mesoscopic RC circuit is represented by
the equivalent circuit of Fig.1B \cite{Buttiker93,Pretre96}. The
geometrical capacitance $C$ is in series with the quantum admittance
$g_{q}(\omega)$ connecting the ac current flowing in the QPC to the
ac internal potential of the dot:
\begin{equation}\label{cond0}
g_{q}(\omega)=\frac{1}{\frac{h}{2e^{2}}+\frac{1}{-i\omega C_{q}}}
\qquad (T = 0)
\end{equation}
The nonlocal quantum impedance behaves as if it were the series
addition of a quantum capacitance $C_{q}$ with a constant contact
resistance $h/2e^2$. $C_{q}=e^{2}\frac{d\mathcal{N}}{d\varepsilon}$
is associated with the local density of state
$\frac{d\mathcal{N}}{d\varepsilon}$ of the mode propagating in the
dot, taken at the Fermi energy. The striking effect of phase
coherence is that the QPC transmission probability $D$ affects the
quantum capacitance (see Eq.\ref{Cq}) but not the resistance. The
total circuit admittance $G$ is simply :
\begin{equation}\label{grandcon}
 G=\frac{-i\omega C g_{q}(\omega)}{-i\omega C+g_{q}(\omega)}= \frac{-i\omega C_{\mu}\frac{2e^{2}}{h}}{-i\omega
 C_{\mu}+\frac{2e^{2}}{h}}\quad , \qquad (T = 0)
\end{equation}
where $C_{\mu}=\frac{CC_{q}}{C+C_{q}}$ is the electrochemical
capacitance. In the incoherent regime, both resistance and quantum
capacitance vary with transmission. The dot forms a second reservoir
and the electrochemical capacitance $C_{\mu}$ is in series with the
QPC resistance $R$. In particular, when the temperature is high
enough to smooth the capacitor density of states, the Landauer
formula $R=\frac{h}{e^{2}}\times\frac{1}{D}$ is recovered.

Several samples have been measured at low temperatures, down to 30
mK, which show analogous features. We present results on two samples
made with 2DEG defined in the same high-mobility GaAsAl/GaAs
heterojunction, with nominal density $n_{s}=1.7 \times 10^{15}$
m$^{-2}$ and mobility $\mu =260$ V$^{-1}$m$^{2}$s$^{-1}$. A finite
magnetic field ($B=1.3$ T) is applied, so as to work in the
ballistic quantum Hall regime with no spin degeneracy \cite{SOM}.

\begin{figure}[hhh]
\centerline{\includegraphics[width=9 cm,
keepaspectratio]{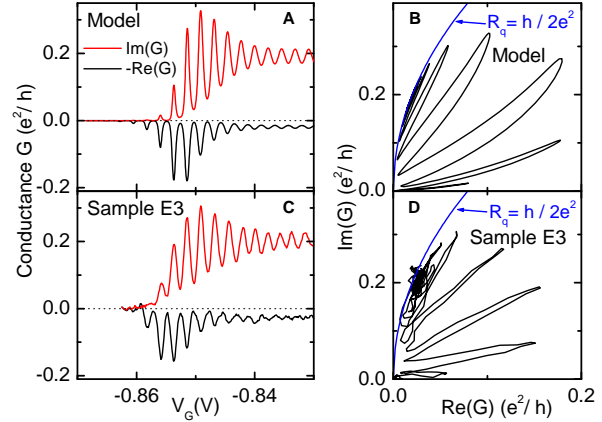}}\caption{{\small Complex conductance of
sample E3 as function of the gate voltage $V_G$ for $T=100$ mK and
$\omega /2\pi=1.2$ GHz, at the opening of the first conduction
channel (C) and its Nyquist representation in (D). The theoretical
circle characteristic of the coherent regime is shown as a solid
line. (A and B) show the corresponding curves for the simulation of
sample E3 using the 1D model with $C=4$ $ fF$, $C_\mu=1$ $
fF$.}}\label{HBT2}\end{figure}

The real and imaginary parts of the admittance $Im(G)$ and $Re(G)$
as a function of QPC gate voltage $V_{G}$ at the opening of the
first conduction channel are shown in Fig.2C. On increasing $V_{G}$,
we can distinguish three regimes. At very negative $V_{G}\leq -0.86$
V, the admittance is zero. Starting from this pinched state, peaks
are observed in both $Im(G)$ and $Re(G)$. Following a maximum in the
oscillations, a third regime occurs where $Im(G)$ oscillates nearly
symmetrically about a plateau while the oscillation amplitude
decreases smoothly. Simultaneously, peaks in $Re(G)$ quickly
disappear to vanish in the noise.

Comparing these observations with the results of
refs.\cite{Buttiker93,Pretre96},using a simplified one-dimensional
(1D) model for $C_q$ with one conduction mode and a constant energy
level spacing in the dot $\Delta$ \cite{Gabelli06}, the simulation
(Fig.2A) shows a striking similarity to the experimental conductance
traces in Fig.2C. In this simulation, $V_{G}$ determines the
transmission $D$ but also controls linearly the 1D dot potential.
The transmission is chosen to vary with $V_G$ according to a
Fermi-Dirac-like dependence appropriate to describe QPC transmission
\cite{Butt90}. This model can be used to get a better understanding
of the different conductance regimes. Denoting $r$ and $t$ the
amplitude reflection and transmission coefficients of the QPC
($r^{2}=1-D$, $t=\sqrt{D}$), we first calculate the scattering
amplitude of the RC circuit:
\begin{equation}\label{S}
s(\varepsilon)=r-t^{2}e^{i\varphi}\sum_{n=0}^{\infty}(re^{i\varphi})^{n}=\frac{r-e^{i\varphi}}{1-re^{i\varphi}}
\end{equation}
where $\varepsilon$ is the Fermi energy relative to the dot
potential and $\varphi=2\pi\varepsilon/\Delta$ is the phase
accumulated for a single turn in the quantum dot. The
zero-temperature quantum capacitance is then given by:
\begin{equation}\label{Cq}
C_q=e^{2}\frac{d\mathcal{N}}{d\varepsilon}=\frac{1}{2i\pi}s^{+}\frac{\partial
s}{\partial
\varepsilon}=\frac{e^{2}}{\Delta}\frac{1-r^{2}}{1-2r\cos2\pi\frac{\varepsilon}{\Delta}+r^{2}}
\end{equation}
Therefore, $C_q$ exhibits oscillations when the dot potential is
varied. When $r \rightarrow 0$,  these oscillations vanish and
$C_q\rightarrow e^{2}/\Delta$. As reflection increases, oscillations
are growing with maxima $\frac{e^{2}}{\Delta} \frac{1+r}{1-r}$ and
minima $\frac{e^{2}}{\Delta} \frac{1-r}{1+r}$. For strong
reflection, Eq \ref{Cq} gives resonant Lorentzian peaks with an
energy width $D\Delta/2$ given by the escape rate of the dot.
However, at finite temperature, the conductance in equation
(\ref{cond0}) has to be thermally averaged to take into account the
finite energy width of the electron source so that :
\begin{eqnarray}\label{condT}
     g_{q}(\omega) = \int d\varepsilon \left ( - \frac{\partial f}{\partial \varepsilon}\right ) \frac{1}{h/2e^2+1/(-i\omega
     C_{q})}\quad \\
     (T\neq 0)  \nonumber
\end{eqnarray}
where $f$ is the Fermi-Dirac distribution. Again, the nonlocal
admittance behaves as if it were the serial association of a
charge-relaxation resistance $R_q$ and a capacitance that we still
denote $C_q$. In the weak transmission regime ($D\rightarrow 0$),
when $D\Delta \ll k_{B}T $, equation (\ref{condT}) yields thermally
broadened capacitance peaks with
\begin{equation}\label{fanC}
C_{q} \simeq
\frac{e^{2}}{4k_{B}T\cosh^{2}(\delta\varepsilon/2k_{B}T)}\quad,\qquad
(D<<1)
\end{equation}
where $\delta \varepsilon$ denotes the energy distance to a resonant
dot level. Note that these capacitance peaks do not depend on the
dot parameters and can be used as a primary thermometer. Similar but
transmission-dependent peaks are predicted in the inverse resistance
\begin{equation}\label{fanR}
1/R_{q} \simeq
D\frac{e^{2}}{h}\frac{\Delta}{4k_{B}T\cosh^{2}(\delta\varepsilon/2k_{B}T)},\qquad
(D<<1)
\end{equation}
This result is reminiscent of the thermally broadened resonant
tunneling conductance for transport through a quantum dot. A
consequence of the finite temperature is the fact that the
resistance is no longer constant. This thermally-induced divergence
of $R_q$ at low transmission restores a frequency-dependent
pinch-off for $R_q\gg 1/C_q\omega$, as can be seen in both model and
experiment in Figs.2A.C. As mentioned above, for $k_{B}T \gg
D\Delta$, the quantum dot looks like a reservoir and the Landauer
formula is recovered.

The coherent and the thermally broadened regimes are best
demonstrated in the Nyquist representation $Im(G)$ versus $Re(G)$ of
the experimental data in Fig.2D. This representation allows to
easily distinguish constant resistance from constant capacitance
regimes, as they correspond to circles respectively centered on the
real and imaginary axis. Whereas, for low transmission, the Nyquist
diagram strongly depends on transmission, the conductance
oscillations observed in Fig.2C collapse on a single curve in the
coherent regime. Moreover this curve is the constant $R_q=h/2e^2$
circle. By contrast, admittance peaks at low transmission correspond
to a series of lobes in the Nyquist diagram, with slopes increasing
with transmission in qualitative agreement with Eqs.\ref{fanC} and
\ref{fanR}. These lobes and the constant $R_q$ regime are well
reproduced by the simulations in Fig.2B. Here, the value of $C_\mu$
and the electronic temperature are deduced from measurement. In our
experimental conditions, the simulated traces are virtually free of
adjustable parameters as $C\geq 4C_\mu \gg C_{q}$.

It is important to note that in a real system, the weak transmission
regime is accompanied by Coulomb blockade effects that are not taken
into account in the above model. In the weak transmission regime and
$T=0$, using an elastic co-tunneling approach
\cite{Averin92,Glattli93}, we have checked that there is no
qualitative change except for the energy scale that now includes the
charging energy so that $\Delta$ is replaced by
$\Delta$+$e^2/C$=$e^2/C_{\mu}$. At large transmission, the problem
is nonperturbative in tunnel coupling and highly nontrivial.
Calculations of the thermodynamic capacitance exist
\cite{Matveev,Glazman98,Brouwer05}, but at present, no comprehensive
model is available that would include both charge-relaxation
resistance and quantum capacitance for finite temperature and/or
large transmission.
\begin{figure}[hhh]
\centerline{\includegraphics[width=3.5 in,
keepaspectratio]{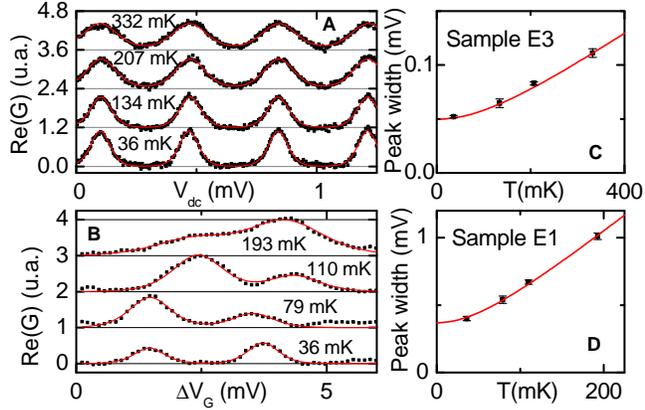}}\caption{{\small Coulomb blockade
oscillations in the real part of the ac conductance in the
low-transmission regime. The control voltage is applied to the
counter-electrode for Sample E3 (A) and to the QPC gate for Sample
E1 (B). The temperature dependence is used for absolute calibration
of our setup, as described in the text : the peak width, shown in (C
and D) as a function of temperature, is deduced from theoretical
fits (solid lines) using Eq.(\ref{fanR}) and taking a linear
dependence of energy with the control voltage. Lines in (C) and (D)
are fits of the experimental results using a $\sqrt{T^2+T_0^2}$ law
to take into account a finite residual electronic temperature
$T_{0}$.}}\label{HBT3}\end{figure} Calibration of our admittance
measurements is a crucial step toward extracting the absolute value
of the constant charge-relaxation resistance. As at GHz frequencies,
direct calibration of the whole detection chain is hardly better
than 3dB, we shall use here an indirect, but absolute, method, often
used in Coulomb blockade spectroscopy, that relies on the comparison
between the gate voltage width of a thermally broadened Coulomb peak
($\propto k_{B}T$) and the Coulomb peak spacing ($\propto
e^2/C_\mu$). From this, an absolute value of $C_{\mu}$ can be
obtained. The real part of the admittance of Sample E3 is shown as a
function of the dc voltage $V_{dc}$ at the counter-electrode, for a
given low transmission (Fig.3A). A series of peaks with periodicity
$\Delta V_{dc}=370\, \mu$V are observed, with the peaks accurately
fitted by Eq.\ref{fanR}. Their width, proportional to the electronic
temperature $T_{el}$, is plotted versus the refrigerator temperature
$T$ (see Fig.3C). When corrected for apparent electron heating
arising from gaussian environmental charge noise, and if we assume
$T_{el}=\sqrt{T^{2}+T_{0}^{2}}$, the energy calibration of the gate
voltage yields $C_\mu$ and the amplitude $1/C_{\mu}\omega$ of the
conductance plateau in Fig.2. A similar analysis is done in Fig.3, B
and D, for sample E1 using $V_G$ to control the dot potential. Here
peaks are distorted because of a transmission-dependent background
and show a larger periodicity $\Delta V_{G}=2$ mV, which reflects
the weaker electrostatic coupling to the 2DEG.

\begin{figure}[hhh]
\centerline{\includegraphics[width=9 cm,
keepaspectratio]{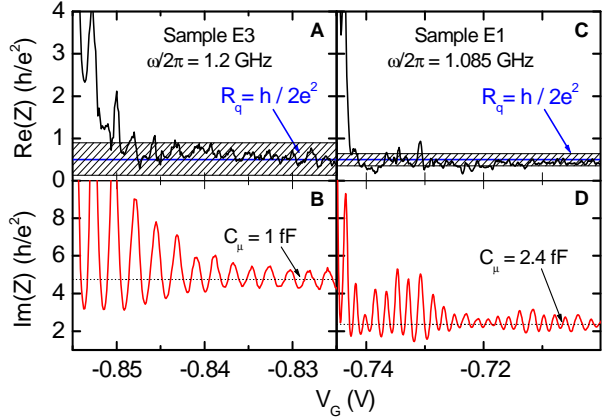}}\caption{{\small Complex impedance of
Sample E3 (A and B) and Sample E1 (C and D) as a function of QPC
voltage for $T= 30$ mK and $B=1.3$ T. The dashed lines in (B and D)
correspond to the values of $C_{\mu}$ deduced from calibration. The
horizontal solid lines in (A and C) indicate the half-quantum of
resistance expected for the coherent regime. Uncertainties on
$R_{q}$ are displayed as hatched areas.}}\label{HBT4}\end{figure}

Finally, after numerical inversion of the conductance data, we can
separate the complex impedance into the contributions of the
capacitance, $1/C_\mu\omega$, and the relaxation resistance $R_q$.
The results in Fig.4 demonstrate deviations from standard
Kirchhoff's laws : the charge-relaxation resistance $R_q$ remains
constant in the regime where the quantum capacitance exhibits strong
transmission-dependent oscillations; this constant value equals,
within experimental uncertainty, half the resistance quantum as
prescribed by theory \cite{Buttiker93,Pretre96}.  In the weak
transmission regime, the Landauer formula is recovered because of
thermal broadening, and $R_q$ diverges as it does in the dc regime.
Furthermore, additional measurements at $4K$ prove that the
classical behavior is indeed recovered in the whole transmission
range whenever $k_BT\gg e^2/C_\mu$.

In conclusion, we have experimentally shown that the series
association of a quantum capacitor and a model quantum resistor
leads to a violation of the dynamical Kirchhoff's law of impedance
addition. In the fully coherent regime, the quantum resistor is no
longer given by the Landauer formula but by the half-quantized
charge-relaxation resistance predicted in
Ref.\cite{Buttiker93,Pretre96}.\\
The LPA is the CNRS-ENS UMR8551 associated with universities Paris 6
and Paris 7. The research has been supported by AC-Nanoscience,
SESAME grants and ANR-05-NANO-028.


\begin{thebibliography}{9}
\bibitem{SOM} Materials and methods are available as supporting material on
Science Online.
\end{thebibliography}
\end{document}